\documentclass[final]{ws-mpla}
\def\draftnote{}%
\usepackage[super]{cite}
\usepackage{graphicx}
\usepackage[breaklinks]{hyperref}  
\hypersetup{colorlinks,urlcolor=black,citecolor=black,linkcolor=black,filecolor=black}
\usepackage{breakurl}
\usepackage{comment}
\usepackage{multirow}

\def\hc{\ensuremath{\mathrm{h.c.}}}
\def\Br{\ensuremath{\mathrm{Br}}}
\def\SM{\ensuremath{\mathrm{SM}}}

\def\GeV{\ensuremath{~\mbox{GeV}}}
\def\MeV{\ensuremath{~\mbox{MeV}}}
\def\cBA{\ensuremath{c_{\beta\alpha}}}
\def\sBA{\ensuremath{s_{\beta\alpha}}}
\def\tB{\ensuremath{t_{\beta}}}
\def\mA{\ensuremath{m_{A}}}
\def\mH{\ensuremath{m_{H}}}

\def\ttH{\emph{ttH}}
\def\VBF{\emph{VBF}}
\def\ZH{\emph{ZH}}

\begin{document}

\markboth{A.~Bednyakov, V.~Rutberg}
{FCNC decays of the Higgs bosons in the BGL model}

\catchline{}{}{}{}{}

\title{FCNC decays of the Higgs Bosons in the BGL model
}

\author{\footnotesize Alexander Bednyakov}

\address{Bogoliubov Laboratory of Theoretical Physics, Joint Institute for Nuclear Research, \\ Dubna, Russia,\\
Dubna State University, Dubna, Russia \\
bednya@theor.jinr.ru}

\author{Veronika Rutberg}

\address{Moscow Institute of Physics and Technology, Dolgoprudny, Russia\\
veronika.rutberg@yandex.ru
}

\maketitle


\begin{abstract}
We consider flavor-changing decays of neutral Higgs bosons in the context of CP-conserving
BGL model --- a variant of 2HDM Type 3 model suggested by Branco, Grimus and Lavoura --- in which tree-level FCNC couplings are suppressed by elements of known fermion mixing matrices.
The relevant regions of parameter space compatible with experimental restrictions on the SM Higgs properties are studied.
We also include current bounds on $h\to \mu\tau$ into consideration. In addition, different FCNC decay modes are analyzed for heavier Higgs states ($H/A$) and conservative estimates for $\Br(A/H\to\mu\tau)$ are provided. 
We updated previous studies and found that it can not be more than 30\%
for heavy Higgses with masses around 350 GeV. 

\end{abstract}

\ccode{PACS Nos.: 12.60.Fr. 14.80.Ec, 14.80.Bn}

\section{Introduction}	

It is a well-known fact that the Higgs field plays a fundamental role in the Standard Model (SM),
giving rise to masses of both the SM gauge bosons and fermions.
The Higgs boson --- the only component of the Higgs doublet, which is not ``eaten'' by massive gauge fields, 
was predicted in 1964. After years of searches it was finally observed at the LHC \cite{Aad:2012tfa,Chatrchyan:2012ufa} in 2012. 
In subsequent years the production and decay processes of the new particle were carefully investigated 
and it was confirmed with rather high precision that there is no significant deviations from the SM predictions.

Nevertheless, in spite of the fact that the SM turns out to be very successful in description of plethora of electroweak phenomena
it is believed that it is not the ultimate theory. 
Among different possibilities to go beyond the SM (BSM) one can consider an extension with an additional Higgs doublet - the so-called 
Two-Higgs-Doublet model (2HDM) (for review see, e.g., Refs.~\refcite{Branco:2011iw,Ivanov:2017dad}). The model being one of the simplest (renormalizable) alternatives of the SM predicts additional scalar
states in the spectrum --- two neutral $A,H$ and one charged $H^\pm$ Higgs bosons. Being (linear combinations of) components of SU(2) doublets
components, its interactions with vector bosons are fixed by the gauge principle but there is a large freedom 
in self-interactions and Yukawa couplings to fermions.

In this paper we consider the so-called Type III 2HDM, in which both doublets couple to up- and down-type fermions 
and, as a consequence, one has tree-level flavor-changing neutral current (FCNC) interactions.
In order to minimize the number of additional parameters were choose a special variant of Type III 2HDM - 
the model proposed by Branco, Grimus and Lavoura (BGL) \cite{Branco:1996bq}, 
in which, due to certain type of symmetry, all the tree-level FCNC interactions\cite{Botella:2015hoa,Alves:2017xmk} are given in terms of known quantities --- fermion masses and mixing matrices, Cabibbo-Kobayashi-Maskawa (CKM) 
in the case of quarks and \& Pontecorvo-Maki-Nakagawa-Sakata (PMNS) in the case of leptons.  

Our main motivation to study such an exotic scenario was initially due to results from ATLAS\cite{Aad:2015gha} and CMS\cite{Khachatryan:2015kon}, 
which indicated that there might 
be a FCNC decay of the observed Higgs boson ($h$) to tau- and mu-leptons\footnote{In what follows, we use shorthand notation $\mu\tau$ to denote the sum $\bar\mu \tau + \bar\tau\mu$.}
, $h\to \mu\tau$. The corresponding branching  fraction $\Br(h\to \mu\tau)$ 
were estimated to be $0.84^{+0.39}_{-0.37}~\%$ by CMS (at 8 TeV with 19.7 fb$^{-1}$). 
The result due to ATLAS was less significant and the best-fit value was $0.77\pm0.62~\%$ corresponding to an integrated luminocity of 20.3 fb$^{-1}$ at  $\sqrt s=$ 8 TeV  .
Unfortunately, subsequent analyses of ATLAS \cite{Aad:2016blu} at 8 TeV and CMS \cite{Sirunyan:2017xzt} at 13 TeV do not confirm the observation of the discussed mode, but just impose constraints on the branching fraction, $\Br(h\to \mu \tau)<1.43\%$  and $\Br(h\to \mu\tau)<0.25\%$ at 95\% CL, respectively.

In the BGL model the decay was studied~\cite{Botella:2015hoa,Sher:2016rhh} and constraints on the model parameter space were found.
Moreover, in the same context the FCNC decay modes of heavy Higgs partners were also investigated in Ref~\refcite{Sher:2016rhh}. 
We updated the previous analysis\cite{Sher:2016rhh} and took new experimental constraints into account. We also found that the $H/A\to cc$ mode, which was not considered in the above-mentioned paper, can substantially reduce other branching ratios for high values of $\tan\beta$.  Our updated analysis gives rise to the upper bound on $\Br(A/H\to \mu\tau)\lesssim 30 \%$.

In order to simplify our study we restrict ourselves to CP-conserving scenario with simple Higgs potential. In this framework, we keep Higgs boson masses as free parameters and concentrate on Yukawa interactions
of the latter.  

The paper organized as follows. In section \ref{sec:bgl} a brief description of the BGL model is provides. In section~\ref{sec:h_constraints}
the constraints on the 
parameter space coming from the experimental bounds on the properties of the lightest, SM-like, Higgs boson
are considered. 
The analysis of FCNC decays of neutral heavy Higgses can be found in section~\ref{sec:fcnc_heavy}.   
The section \ref{sec:conclusions} is devoted to conclusions.

\section{\label{sec:bgl} BGL model}
In general 2HDM of Type 3 the Yukawa interaction Lagrangian can be cast into the following form
\begin{eqnarray}
	\mathcal{L}_{Y}& = & -\bar{Q} (\Phi_{1} y^{d}_{1} + \Phi_{2} y^{d}_{2}) {d}_{R}-\bar{Q}(\tilde{\Phi}_{1} y^{u}_{1} + \tilde{\Phi}_{2} y^{u}_{2}) u_{R}   \nonumber\\
		       & &  - \bar{L} (\Phi_{1} y^{l}_{1} + \Phi_{2} y^{l}_{2}) {l}_{R}-\bar{L} (\tilde{\Phi}_{1} y^{\nu}_{1} + \tilde{\Phi}_{2} y^{\nu}_{2}) \nu_{R} + \hc,
\end{eqnarray}
	in which the summation over generations is implied. 
Both Higgs doublets $\Phi_i$ 
\begin{eqnarray}
	\Phi_i = \begin{pmatrix}
    			\phi^+_i \\
                \frac{v_i + \phi_i^0 + i \eta_i^0}{\sqrt 2} 
			\end{pmatrix}
            ,
                \qquad \tan\beta \equiv v_2/v_1, 
                \quad v = \sqrt{v_1^2 + v_2^2} \simeq 246~\GeV 
                \label{eq:Gen_2HDM}
\end{eqnarray}
	couple to right-handed (RH) fermions 
	-- down-type quarks $d_R$ and charged leptons $l_R$. The corresponding charge-conjugated doublets $\tilde \Phi_i\equiv - i \sigma_2 \Phi^*_i$ couple to RH up-type fermions\footnote{We assume here that neutrinos are Dirac fermions. BGL models with Majorana neutrino are discussed, e.g., in Ref. \refcite{Botella:2011ne}.} 
$u_R$, $\nu_R$.
	The left-handed (LH) SU(2) doublets for quarks and leptons  are denoted in \eqref{eq:Gen_2HDM} by $Q$ and $L$, respectively. 
    
    The flavor-changing neutral current interactions 
    appear due to the fact that Yukawa matrices $y_i^{f}$  in flavor space are not diagonalizable simultaneously for $i=1,2$ and fixed $f=\{u,d,l,\nu\}$. In general, the size of FCNC interactions can be arbitrary large and is constrained only by experiment. An attractive  feature of the so-called  
    Branco, Grimus and Lavoura (or BGL) model \cite{Branco:1996bq} is that FCNC in the Higgs sector are related to the parameters of fermion mixing matrices, i.e. CKM ($V_{ij}$) - for the quark sector and PMNS ($U_{ij}$ )- in the case of leptons.   
   
    Branco, Grimus and Lavoura demonstrated that if the Yukawa term in the Lagrangian  respects a symmetry for a generation number $k$ 
   \begin{equation}
Q^{k} \to e^{i \tau} Q^{k},\qquad
u^{k}_{R} \to e^{i 2\tau} u^{k}_{R}, \qquad 
\Phi_{2} \to e^{i \tau} \Phi_{2}, 
\qquad \tau \neq 0,\frac{\pi}{2}, \label{eq:BGLsym:up}
\end{equation} or
\begin{equation}
Q^{k} \to e^{i \tau} Q^{k},\qquad
d^{k}_{R} \to e^{i 2\tau} d^{k}_{R}, \qquad 
\Phi_{2} \to e^{-i \tau} \Phi_{2}, 
\qquad \tau \neq 0,\frac{\pi}{2}, \label{eq:BGLsym:down}
\end{equation}
the couplings of fermionic decays can be expressed through the elements of mixing matrix $U$ (or $V$, when Eqs.\eqref{eq:BGLsym:up} and \eqref{eq:BGLsym:down} are generalized to include leptons). 
In addition, it can be shown (Ref. \refcite{Branco:1996bq}) that if the symmetry are imposed on the Yukawa term for up-type fermions \eqref{eq:BGLsym:up}, FCNC will appear only in processes with down-type fermions, and vice versa. In this way, in leptonic sector only up-type BGL models are interesting for investigation of the decays  $h/H/A \rightarrow \mu \tau$.  
If the symmetry is imposed on the Yukawa term for down-type quarks \eqref{eq:BGLsym:down}, the FCNC processes with $t$-quark can be observed with large probabilities. 
 We do not want to consider such scenarios here and restrict ourselves only to up-type BGL models for both leptons and quarks.

 The generation number $k$ is to be chosen for leptons in such a way that heavy Higgs bosons decay widths $\Gamma(H \rightarrow \mu \tau)$ and $\Gamma(A \rightarrow \mu \tau)$ can be significantly enhanced. The tree-level width of the processes $H/A \rightarrow \mu \tau$ in $k$-generation up-type BGL is given by \cite{Botella:2015hoa}
\begin{align}
\Gamma(H \rightarrow \mu \tau )=\frac{m_{H}m^{2}_{\tau}}{8 \pi v^{2}}(\tan \beta + \cot \beta)^{2}|U^{*}_{\tau k} U_{\mu k}|^{2} \sin^{2} (\beta - \alpha) \label{eq:Gamma_H}, \\
\Gamma(A \rightarrow \mu \tau )=\frac{m_{A}m^{2}_{\tau}}{8 \pi v^{2}}(\tan \beta + \cot \beta)^{2}|U^{*}_{\tau k}U_{\mu k}|^{2} \label{eq:Gamma_A}.
\end{align}
where we neglect all lepton masses but that of tau $m_\tau$.
Both equations~\eqref{eq:Gamma_H} and \eqref{eq:Gamma_A} depend 
on $v^2$ and $\tB \equiv \tan\beta$. 
In addition, the mixing angle 
$\beta-\alpha$ corresponding to rotation in the CP-even Higgs sector from  
Higgs to mass basis enters Eq.~\eqref{eq:Gamma_H}.
In what follows, instead of $\beta-\alpha$ we utilize $\cBA\equiv \cos (\beta-\alpha)$ with $0 \leq \beta - \alpha \leq \pi$, which parametrize the deviation of the lightest CP-even boson $h$ from that of the SM (c.f., Ref.\refcite{Haber:2015pua}).

 Substituting numerical values of the PMNS matrix elements into the factors $|U^{*}_{ \tau k}U_{\mu k }|^{2}$ we have
\begin{align}
|U^{*}_{ \tau 1}U_{\mu 1 }|^{2}
:
|U^{*}_{ \tau 2}U_{\mu 2 }|^{2}
:
|U^{*}_{ \tau 3}U_{\mu 3 }|^{2}
\simeq  
(2:12:24) \times 10^{-2},
 \end{align}
 so one can see that the largest value corresponds to $k=3$. Due to this, in what follows we use $k=3$ in the case of leptons. 
 In spite of the fact that the BGL generation $k$ for quarks ($k_q$) in \eqref{eq:BGLsym:up} can be chosen different from the leptonic one ($k_l$), we assume $k_q=k_l=k=3$ in anticipation of quark-lepton symmetry.

It is worth mentioning, however, that due to hierarchical structure of the CKM matrix, in the BGL model of quark type $k$ 
the diagonal couplings of heavy Higgses to quarks exhibit similar 
pattern both for up- and down-type quarks and scale as $\tB$ or
$1/\tB$ for $A$-boson interacting with $i\neq k$ or $k$-quark generation. 
For $H$-boson the dependence is similar near $\cBA=0$ (decoupling/alignment limit\cite{Gunion:2002zf}), but due to mixing in the CP-even sector becomes 
more involved away from this line.

Concerning 2HDM model one should keep in mind that general Higgs potential (see,e.g., Ref.~\refcite{Branco:2011iw}) 
\begin{align}
	V_{Higgs} &  =m^{2}_{11}\Phi^{\dag}_{1}\Phi_{1}+m^{2}_{22}\Phi^{\dag}_{2}\Phi_{2}-(m^{2}_{12}\Phi^{\dag}_{1}\Phi_{2}+~h.c.~) \nonumber \\
	& +\frac{\lambda_{1}}{2}(\Phi^{\dag}_{1}\Phi_{1})^{2}+\frac{\lambda_{2}}{2}(\Phi^{\dag}_{2}\Phi_{2})^{2} + \lambda_{3}\Phi^{\dag}_{1}\Phi_{1}\Phi^{\dag}_{2}\Phi_{2}+\lambda_{4}\Phi^{\dag}_{1}\Phi_{2}\Phi^{\dag}_{2}\Phi_{1} \nonumber\\ 
	& +\left[\frac{\lambda_{5}}{2}(\Phi^{\dag}_{1}\Phi_{2})^{2}+\lambda_{6}\Phi^{\dag}_{1}\Phi_{1}\Phi^{\dag}_{1}\Phi_{2}+\lambda_{7}\Phi^{\dag}_{2}\Phi_{2}\Phi^{\dag}_{1}\Phi_{2}+\hc~\right]
\end{align}
has very rich structure and, so that, it can significantly affect the branching fractions and the space of permissible parameters. In a rigorous analysis for $H$ and $A$ heavier than $2 m_{h}$ with $m_h$ being the lightest Higgs $h$ mass, some additional processes should be involved in the speculation, e.g., $H \to hh$, $H \to AA$ and $H \to H^{\pm} H^{\mp}$.

It turns out that $m_{12}^2$ and $\lambda_{5-7}$ violate the BGL symmetry. 
In our simplified setup we assume that $\lambda_{5-7}=0$ but keep $m_{12}^2\neq 0$ (soft symmetry breaking \cite{Branco:1996bq}). 
The remaining Higgs potential parameters can be traded for the following convenient set (see, e.g., Ref. \refcite{Kanemura:2004mg}): 
Higgs masses $m_h$, $m_H$,  $m_{H^+}$, $m_A^2 = M^2=\frac{m^2_{12}}{c_{\beta} s_{\beta}}$, vacuum expectation value $v$ and
two dimensionless quantities $\tB$ and $\cBA$ mentioned earlier.

In order to carry out our analysis, we routinely utilize a modified version of the \texttt{2HDMC} package\cite{Eriksson:2009ws}.  The code \texttt{2HDMC} allows one to apply different constraints on the Higgs potential (e.g., tree-level stability and unitarity\cite{Ivanov:2017dad}). Moreover, it can be interfaced with \texttt{HiggsBounds 4.3.1}\cite{Bechtle:2013wla} and \texttt{HiggsSignals 2.1.0}\cite{Bechtle:2013xfa} to confront the model with known experimental bounds. 
In addition, we made use of a private \texttt{Mathematica} routine to cross-check the calculation. 

Our choice of parameters is mainly motivated by~Ref.\refcite{Sher:2016rhh}, in which the benchmark masses $m_A=350$ GeV and $m_H=350$ GeV were considered to avoid large branchings to the top-antitop pairs. We performed a scan over the heavy Higgs masses $m_A,m_H,m_H^\pm$ in the range  $320-400$ GeV and found that the most interesting scenarios, which both survive the considered experimental tests and have rather large FCNC branchings, lie in the region of almost degenerate\footnote{The case with $M=m_A=m_H^\pm$ corresponds to $\lambda_4=0$ in the considered model.} masses $m_A=m_H=m_H^\pm$ slightly above the top-pair threshold. Due to this, in what follows we restrict ourselves to the case $m_A=m_H=m_H^\pm=350$ GeV and study the allowed regions in the $\cBA-\tB$ plane.

\section{\label{sec:h_constraints}Constraints from the Properties of Lightest Higgs}

Given the model at hand one can evaluate the total width and the branching fractions of the 
Higgses.
Identifying the lightest state  with the Higgs boson $h$ observed at the LHC\cite{Aad:2012tfa,Chatrchyan:2012ufa} one imposes important constraints on the  Higgs sector of the considered model. 
In this paper we consider constraints on the total width of $h$, on the branching fraction $\Br(h\to \mu \tau)$, and, finally, on the SM Higgs coupling-strength 
modifiers\cite{Khachatryan:2016vau}. The corresponding allowed region can be visualized in the $\cBA-\tB$ plane.

The calculation of the total width $\Gamma_h$ is straightforward. However, it is important to emphasize that we allow deviation from the SM prediction
$\Gamma^{\mathrm{SM}}_h = 4.07$ MeV (see, e.g., Ref.~\refcite{Patrignani:2016xqp}) and compute branching fractions taking it into account. 
For convenience, in the following figures we indicate the contour corresponding to the experimental bounds (at 95 \% CL) on the width:  $\Gamma_h < 13$ MeV \cite{Khachatryan:2016ctc}. 
Since we are interested in FCNC Higgs decays, we tried to account for 
the 95 \% CL bound on $\Br(h\to \mu\tau)$, which we took to be 0.25\% from recent Ref.~\refcite{Sirunyan:2017xzt}.

On the other hand, it is necessary to compare the predictions of the model with some other experimental results.
Since no significant deviation from the SM results are observed, one can exploit a recent analysis 
of the Higgs production and decays (Ref. \refcite{Khachatryan:2016vau}), in which signal strength parameters $\mu$ were introduced to account for possible New Physics contributions to the cross section of the 
process $i \to h \to f$, characterized in the narrow-width approximation by the product $\sigma_i \cdot \Br^f$:
\begin{align}
	\mu_i\equiv \frac{\sigma_i}{(\sigma_i)_{\mathrm{SM}}},
	\qquad
	\mu^f \equiv \frac{\Br^f}{(\Br^f)_{\mathrm{SM}}}.
\end{align}
Motivated by $\kappa$-framework \cite{Heinemeyer:2013tqa}, various signal strength can be rewritten in terms of SM coupling modifiers $\kappa_p$,  
which correspond to rescaling of the SM coupling of particle $p$ to the SM-like Higgs boson (see Table~\ref{tab:coupling_modifiers} for the case of the considered BGL model
). 

\begin{table}[h]
	\tbl{The SM Higgs coupling modifiers in the BGL model with $k=3$ both in quark and lepton sectors. 
		The diagonal couplings to down-type fermions involve elements of CKM, $V_{ij}$, and PMNS, $U_{ij}$, matrices. In addition, the expression for effective coupling modifiers for gluons and photons are provided.}%
		{\begin{tabular}{|l|l|l|l|}
		 \toprule
		   $\kappa_W$ & $\sBA$ & $\kappa_Z$ & $\sBA$ \\
		   $\kappa_{u,c}$ & $\sBA + \tB \cBA$ & $\kappa_t$ & $\sBA - \tB^{-1} \cBA$ \\
		   $\kappa_{d_i}$   &  $\sBA + ( \tB - (\tB + \tB^{-1}) |V_{td_i}|^2  ) \cBA$
		   & $\kappa_{l_j}$ &  $\sBA + ( \tB - (\tB + \tB^{-1}) |U_{\tau j}|^2  ) \cBA$ \\ 
		   \hline
		   $\kappa^2_g$        & $ 1.06 \kappa_t^2 + 0.01 \kappa_b^2 - 0.07 \kappa_t \kappa_b$ 
		   & $\kappa^2_\gamma$ & $ 1.59 \kappa_W^2 + 0.07 \kappa_t^2 - 0.66 \kappa_W \kappa_t$  \\
		   \hline
	   \end{tabular} \label{tab:coupling_modifiers}}
\end{table}

In our analysis we made use of parameterization in terms of ratios of coupling modifiers (see Section  4.2 of Ref.~\refcite{Khachatryan:2016vau})
and introduced the following vector
\begin{align}
	X & = \{\kappa_{gZ}, \lambda_{Zg}, \lambda_{tg}, \lambda_{WZ}, \lambda_{\gamma Z}, \lambda_{\tau Z}, \lambda_{b Z} \},
\label{eq:couplingratios}
\end{align}
	in which
\begin{align}
	\kappa_{gZ} \equiv \kappa_{g} \cdot \kappa_{Z}/\kappa_{h},
	\label{eq:kappagZ}
\end{align}
	with $\kappa^2_h = \sum_f (\Br^f)_{\SM} \kappa_f^2$.
	Eq.~\eqref{eq:kappagZ} describes 
	how the SM $gg\to h\to ZZ$ process is modified due to new values of couplings
	of the SM particles.  The parameters 
\begin{align}
	\lambda_{Zg} & = \kappa_Z/\kappa_g ,\qquad \lambda_{tg} = \kappa_t/\kappa_g,
		\label{eq:lambdasProd} \\
	\lambda_{WZ} & = \kappa_W/\kappa_Z, \qquad \lambda_{\tau Z} = \kappa_{\tau}/\kappa_Z, 
	\qquad \lambda_{b Z} = \kappa_b/\kappa_Z, \qquad \lambda_{\gamma Z} = \kappa_\gamma/\kappa_Z.
		\label{eq:lambdasDec}
\end{align}
 	probe \VBF, \ZH, and \ttH  production channels \eqref{eq:lambdasProd} together with different Higgs decay modes \eqref{eq:lambdasDec}.

In order to apply the constraints on our parameter space, we calculate 
\begin{align}
	\chi^2(\cBA,\tB) = \sum_{ij} \left[\frac{X_i(\cBA,\tB) - X_i^{\mathrm{exp}}}{\delta X_i^{\mathrm{exp}}}\right] 
	C_{ij}^{-1} \left[\frac{X_j(\cBA,\tB) - X_j^{\mathrm{exp}}}{\delta X_j^{\mathrm{exp}}}\right],
\label{eq:chi2}
\end{align}
where $X_i(\cBA,\tB)$ correspond to our predictions in BGL, while $X^{\mathrm{exp}}_i\pm\delta X_i^{\mathrm{exp}}$ and $C_{ij}$ are the best-fit values and the correlation matrix 
for the parameters $X_i$ given in Table 10 and Figure 29 of Ref.\refcite{Khachatryan:2016vau}, respectively.

\begin{figure}[th]
	\centering{\includegraphics[width=0.7\textwidth]{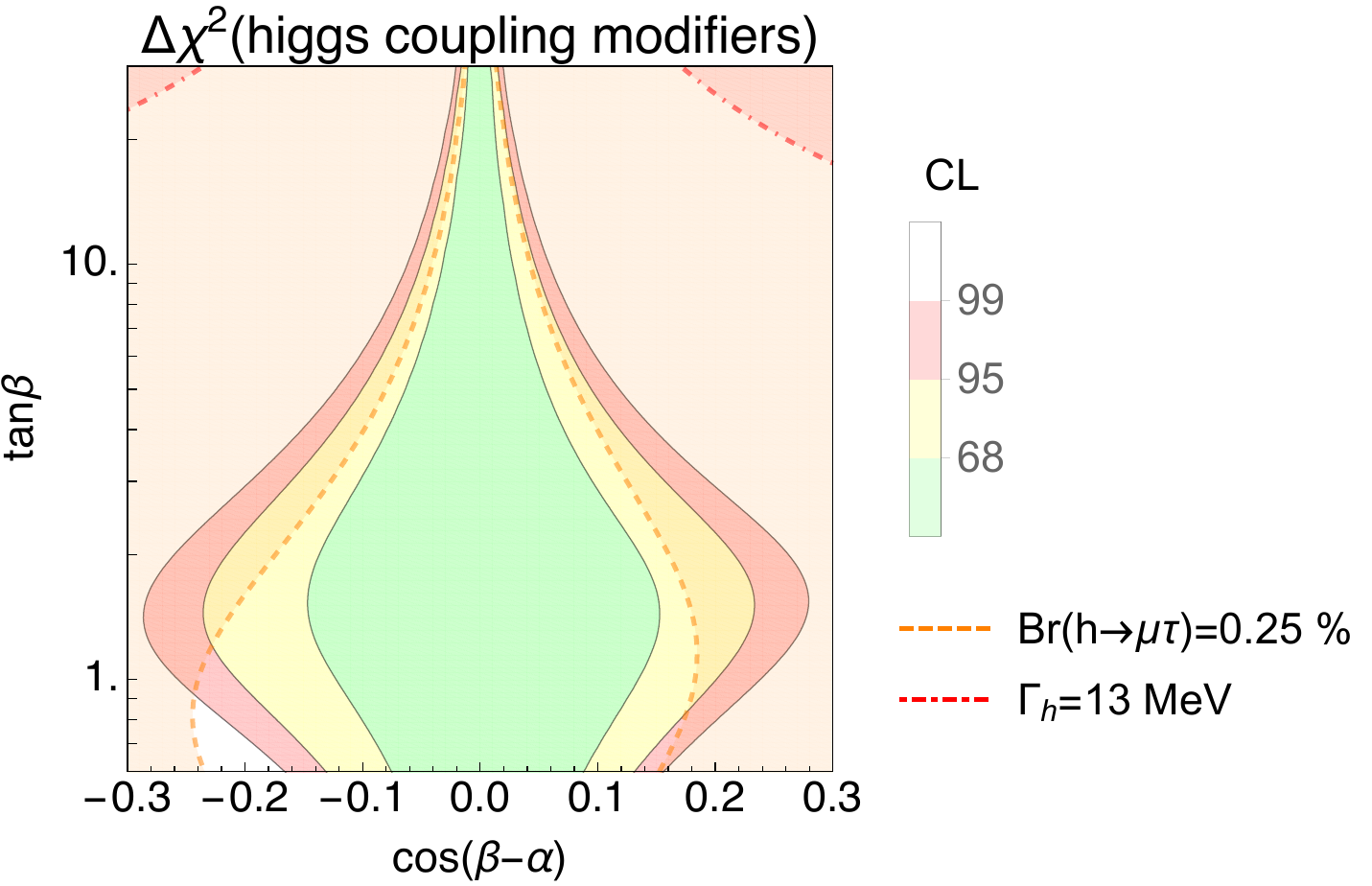}}
	\caption{Constraints on the parameter space due to the lightest Higgs couplings\cite{Khachatryan:2016vau}. Possible deviations from the SM predictions are quantified by 68, 95, and 99 \% CL regions (see. Sec.~\ref{sec:h_constraints}). In addition, regions bounded by the contours corresponding to 95 \% CL limits on the total width\cite{Khachatryan:2016ctc} $\Gamma_h = 13$ MeV and branching\cite{Sirunyan:2017xzt} $\Br(h\to\mu\tau)=0.25\%$ are presented (the allowed area is inside the contours).}
\label{fig:h_constraints}
\end{figure}

It is known from previous studies (see, e.g. Refs.~\refcite{Craig:2013hca,Haber:2015pua}) that experimental data prefers SM-like scenario with $\cBA \to 0$%
. 
In order to quantify possible deviations from this case we assume that $\Delta \chi^2\equiv \chi^2(\cBA,\tB) - \chi_{\mathrm{min}}^2$ 
follows $\chi^2$-distribution with 2 degrees of freedom and 
plot the 68, 95 and 99 \% CL regions 
in the $(\cBA,\tB)$ parameter space (see Fig.~\ref{fig:h_constraints}). 
Due to the fact that the decays to heavy Higgses is kinematically forbidden for $h$,  to large extent the presented analysis is independent of the details of the Higgs potential. 
On can see that the bound $\Br(h\to \mu\tau)$ correlates with constraints on coupling-strength modifiers. In what follows we use the former in combination of other restrictions on the parameter space to single out the allowed region. 

\section{\label{sec:fcnc_heavy}Constraints on FCNC-Decays of Heavy Neutral Higgses}

Non-diagonal parts of Yukawa couplings to heavy neutral Higgses are proportional to $\sin (\beta-\alpha)$ in this model. And it means that in the area, where 2HDM BGL does not deviate from SM too much (near the $\cBA =0$), the heavy Higgses, if they exist, might be detected by their FCNC decays. 
In view of recent experimental results on $h\to \mu\tau$, this is our 
main motivation to study the decays of heavy Higgses.

Before speaking about the decay widths and branching fractions, we should highlight that particles like heavy neutral Higgses have not been observed yet for masses less then 1 TeV (and more then 126 GeV).
This non-observation impose strong constraints on our parameter space.
We utilize the \texttt{HiggsBounds 4.3.1} package\cite{Bechtle:2013wla} to apply such kind of constraints. In addition, we consider recent Refs. \refcite{Aaboud:2016hmk} and \refcite{Aaboud:2016tru,ATLAS:2016eeo,Khachatryan:2016hje,Aaboud:2017yyg} and demand that the cross sections $\sigma(pp\to X  \to \mu \tau)$ and
$\sigma(pp\to X  \to \gamma\gamma)$ to be not higher than 
50 and 5 fb at 13 TeV, respectively, for 
 a new scalar state $X=H/A$. 
For our study we assume that the dominant channel of Higgs productions is the gluon fusion and approximate the considered cross sections at c.m.s. energy $s$ by means of
\begin{align}
	\sigma(pp \to X \to f) & = K \cdot 
	\underbrace{\frac{\Gamma(X \to gg)}{M_X} 
	\cdot \frac{C_{gg}(s,M_X)}{s}}_{\sigma(pp\to X)} \cdot \Br(X \to f) 
\end{align}
	with $f=\gamma\gamma,\mu\tau$, etc. denoting final-state particles, to which a scalar state $X$ decays, and   
\begin{align}
	C_{gg}(s,M_X) & = \int\limits_{M_X^2/s}^1 \frac{d x}{x} g(x) g \Big( \frac{M_X^2}{sx}\Big) 
\end{align}
being dimensionless partonic integral of gluon PDFs $g(x)$.  To evaluate production cross section we use \texttt{ManeParse} package \cite{Clark:2016jgm} and MSTW2008NLO\cite{Martin:2009iq} set of PDFs with $\mu=M_X$.
We also introduce a K-factor $K=1.6$ (see e.g., Ref.~\refcite{Djouadi:2005gi}) to account for QCD corrections at $s=13$ TeV.

\begin{figure}[th]
	\includegraphics[width=1\textwidth]{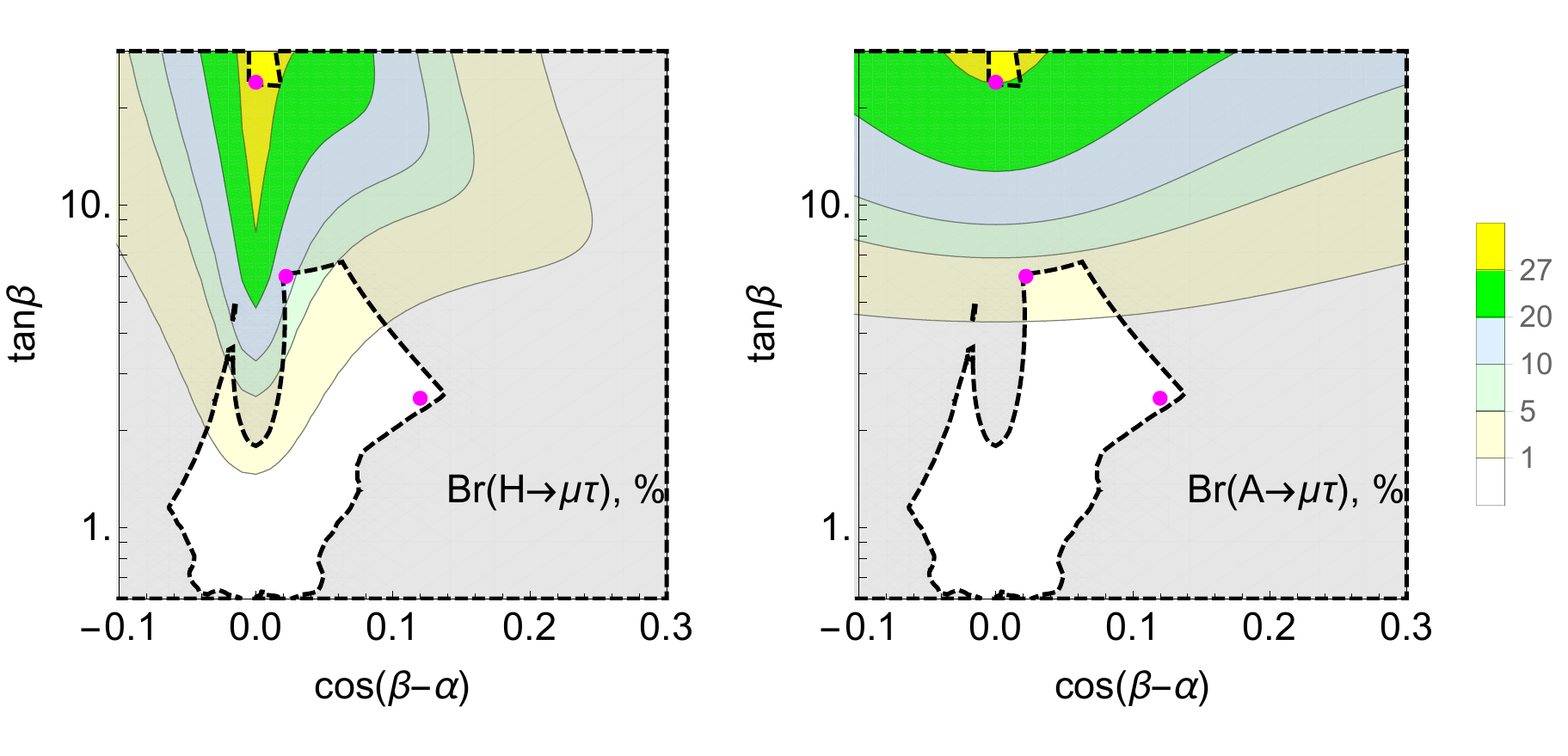}
	\caption{Pictures shows $\Br(H/A \to \mu \tau)$  and all considered constraints (see Fig. \ref{fig:constatints} for details). The case $m_{A}=M=m_{H^{\pm}}=m_{H}=350\GeV$ is presented. Benchmark points discussed in the text are also indicated by purple dots.}
\label{A350_H350}
\end{figure}

\begin{figure}[th]
	\centering{\includegraphics[width=0.7\textwidth]{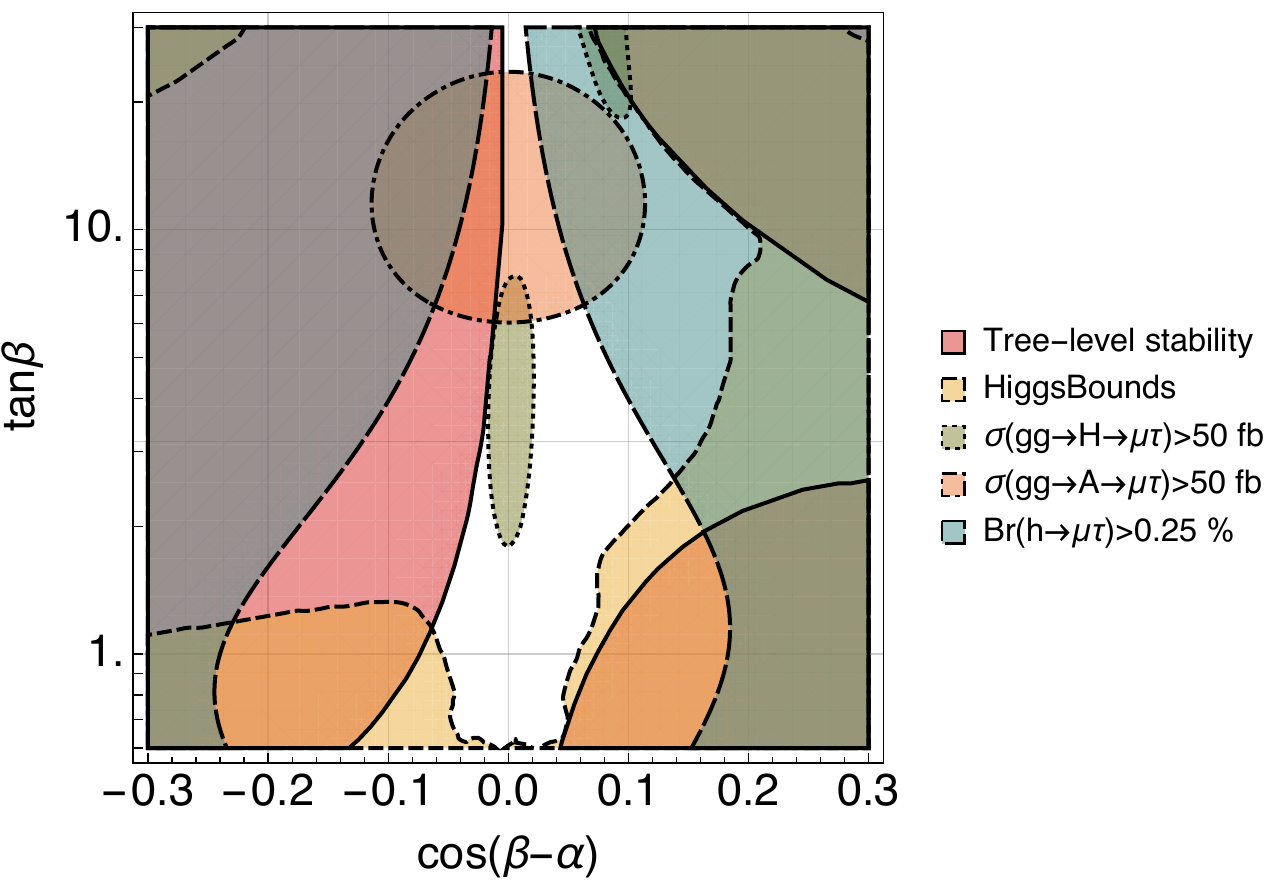}}
\caption{Various constraints on the parameter  space for the case $m_A=m_H=m_{H^+}=350$ GeV.} 
\label{fig:constatints}
\end{figure}

We now turn to branching fractions of heavy Higgses. 
In Fig.~\ref{A350_H350}, we consider the case $m_A=m_H=m_{H^+}=350$ GeV and show $\Br(H/A \to \mu \tau)$ 
together with the allowed region due to all the considered constraints. The latter are also presented in more details in Fig.~\ref{fig:constatints}.

Let us give a few remarks regarding the pictures.  
We consider 
theoretical constraints on the Higgs potentials. 
The crucial condition for stability is that the potential must be bounded from below. It imposes strong restrictions on the allowed parameter space\cite{Branco:2011iw}. At tree level\footnote{We do not utilize here a 
renormalization-group analysis and rely on the output of \texttt{2HDMC}.} we have 
	\begin{align*}
		\lambda_{1}\geq 0, \qquad \lambda_{2}\geq 0,\qquad
		\lambda_{3}\geq -\sqrt{\lambda_{1} \lambda_{2}},\qquad
		\lambda_{3}+\lambda_{4}+|\lambda_{5}|\geq -\sqrt{\lambda_{1} \lambda_{2}}.
	\end{align*}
	From Fig.~\ref{fig:constatints} one can see that for the considered scenario it significantly reduce the allowed region given in Fig.~\ref{fig:h_constraints}.
	In addition, following Ref.~\refcite{Barroso:2012mj} we consider a discriminant $D=(m_{11}^2 - k^2 m_{22}^2)(\tB - k)$ with $k^2=\sqrt{\lambda_1/\lambda_2}$ to check that all the points allowed by the stability constraint correspond to the global minimum ($D>0$). We also check tree-level unitarity (see Appendix A. of Ref.~\refcite{Branco:2011iw}
and the 2HDMC manual \cite{Eriksson:2009ws}).
It turns out that in our case 
the corresponding forbidden region lies completely within the area, in which tree-level stability 
constraint is violated. Due to this, we do not show it in Fig.\ref{fig:constatints}.

From the figure one can deduce that the dependence of the $\sigma(pp\to H/A  \to \mu \tau)$ region boundaries on $\cBA$ is weaker for the case of CP-odd Higgs (only $A \to h Z$ involve $\cBA$ in the considered approximation), while it is more pronounced for the case of $H$-boson. For low $\tB$ the corresponding regions (and those due to the $\gamma\gamma$ final state) are not shown, since they are completely included in the area forbidden by \texttt{HiggsBounds 4.3.1}. 
For large values we have spots with $\sigma(pp\to H/A  \to \mu \tau)$ exceeding the limit. In this situation the suppression of the top-quark contribution (near $\cBA=0$, we have $Y_{ttH/A} \sim \frac{1}{\tB}$)
to $\Gamma(H\to gg)$  is compensated by the enhancement of $\Br(H\to \mu\tau)$.

For the case $m_{A/H} < 2m_t$ the forbidden regions due to $\sigma(gg\to A/H\to \mu\tau/\gamma\gamma)$ obviously enlarge, since below the top-quark threshold, the decay into  top quarks is kinematically forbidden\footnote{The code \texttt{2HDMC} can also account for the decay modes $H/A\to t t^*$ with one off-shell $t$-quark.} and other branchings become large. 
It turns out that already for $m_A=330$ GeV the forbidden area covers almost all the considered parameter space in the $(\cBA,\tB)$-plane.
On the contrary, above the threshold the mode turns on and increases the total width. This leads to suppression of other branchings, especially for small $\tB$. 
For high $\tB$ the decay $H/A\to cc$ is enhanced, so near $\cBA=0$ it can dominate the total width.

\begin{figure}[th]
	\begin{center}
\begin{tabular}{cc}
	\includegraphics[width=0.45\textwidth]{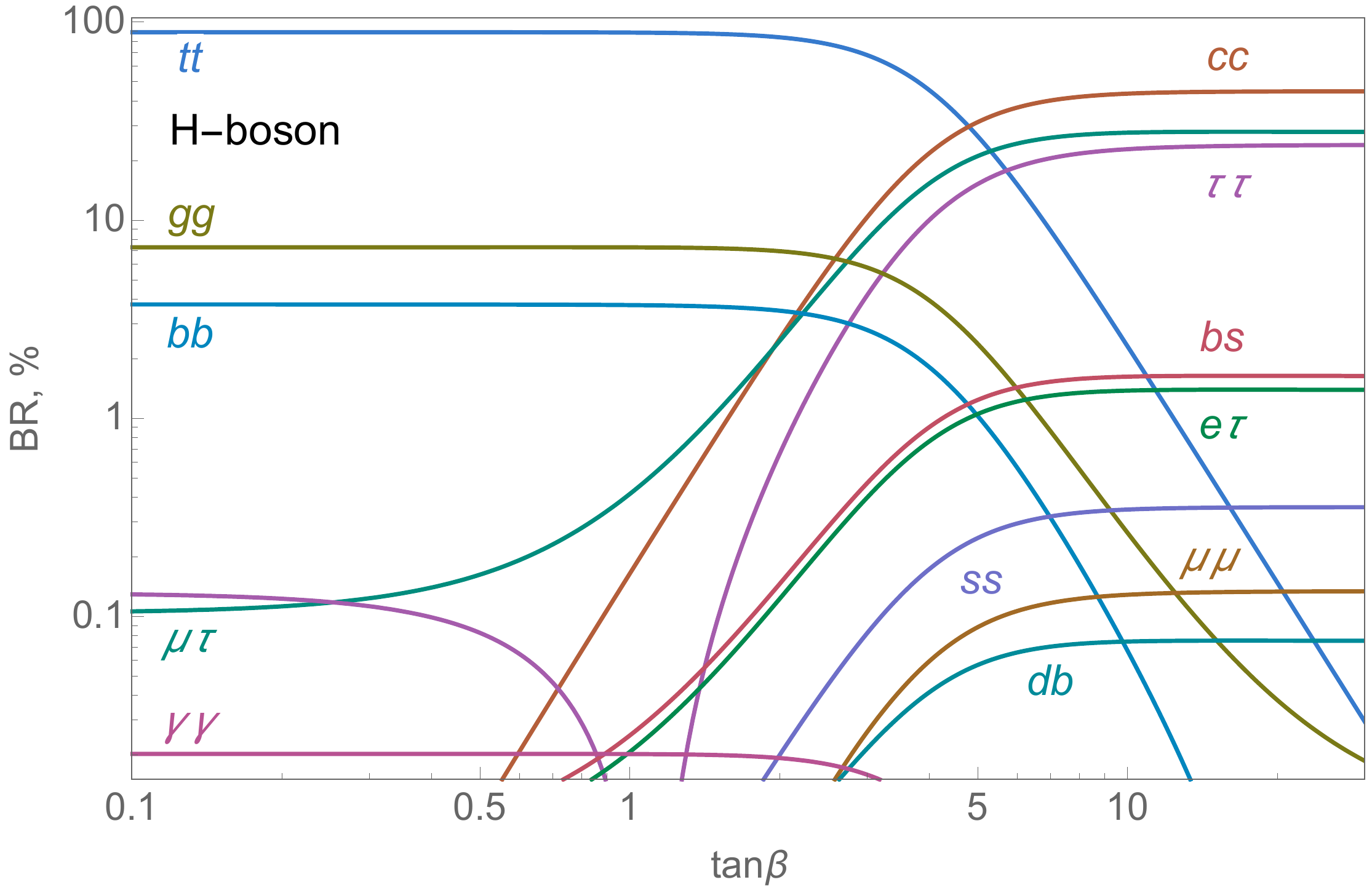}
	&
	\includegraphics[width=0.45\textwidth]{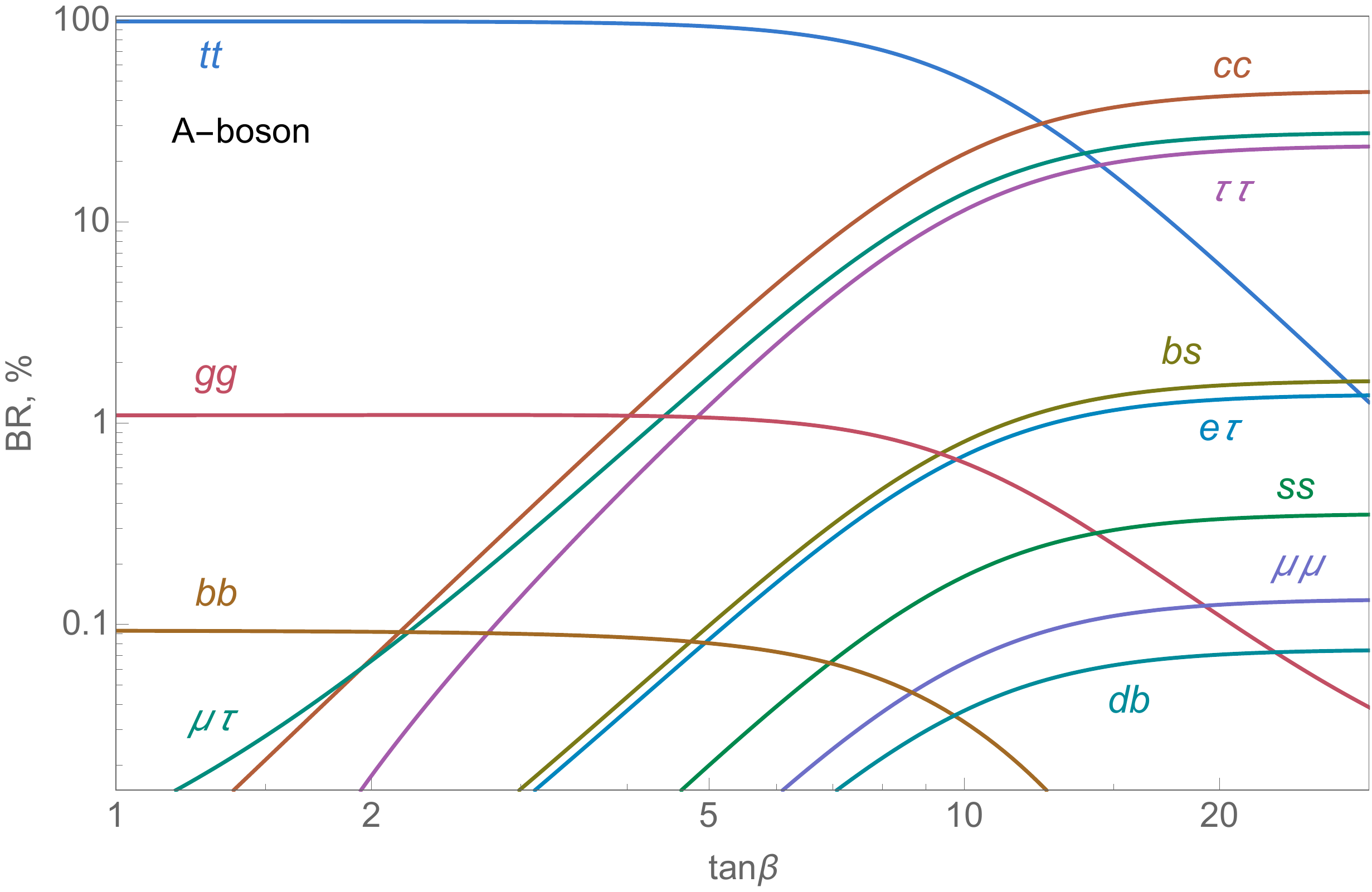}
\end{tabular}
\end{center}
\caption{Various branching fractions of heavy neutral Higgs bosons as functions of $\tB$ for $\cBA=0$ and $m_{A}=M=m_{H^{\pm}}=m_H=
350\GeV$. 
The decays with $\Br<0.01\%$ are not shown. At low values the $t\bar t$ mode dominates, while for high $\tan\beta$ the $c\bar c$ mode becomes important. One can also notice the suppression of the gluon-gluon (gg) mode for large values of $\tan\beta$.} 
\label{fig:br_AH}
\end{figure}

This fact motivates our choice of the heavy Higgs masses slightly above the top threshold ($m=350 ~ \GeV$). In this case, the area near $\cBA=0$ (corresponding to the SM-like scenarios we are interested in) is allowed and $\Br(H/A \to \mu \tau)$ as large as it may be. Moreover, near the threshold the $t \bar t$-mode is suppressed due to kinematics. 
Nevertheless, it is worth emphasizing here the role of $t\bar t$ and $c\bar c$ modes in our analysis of the case $\mH=\mA=350$ GeV.  In Fig.\ref{fig:br_AH} one can see the dependence of different branching fractions of $H$ and $A$ 
on $\tB$ for $\cBA=0$. 
One can see that with the increase of $\tB$ the $c\bar c$-mode becomes the dominant one, while both branchings to $t\bar t$ and $gg$ go down. Let us mention here that  the scenarios with $m_A,m_H=350~\GeV$ were also considered in Refs.~\refcite{Sher:2016rhh}. However, the authors of this paper did not take into account the $t\bar t$ and $c\bar c$ modes\footnote{We thank M.Sher for clarifying us this subtlety.}, and, thus, predicted larger values for $\Br(H/A\to \mu\tau)$.

Regarding the search of heavy Higgses, it is clear that the most interesting points lie near the boundaries of the allowed area, where some of branching fractions and cross sections can be as large as possible. 
The vicinity of $\cBA=0$ is favoured (see the benchmark point BMP1 in what follows), since on this line the maxima of $\Br(A \to \mu \tau)$ and $\Br(H \to \mu \tau)$ are achieved. The experimental bounds can be evaded by choosing rather large $\tB$. 
In addition, we are also interested in scenarios, which can lead to other decays of heavy Higgses observable in the near future. 

For example, if $\tB$ does not deviate much from one, $\cBA$ can substantially differ from zero. 
In this case the modes $A \rightarrow hZ $ and $H \rightarrow hh$ which scale as $\sim \cBA^{2}$ can be enhanced (see BMP2, BMP3\footnote{It is worth pointing that BMP3 also  
features relatively large $\Br(h\to\mu\tau)=0.17\%$.}
). 
In addition, due to peculiarities of the PMNS matrix 
the mode $H /A \rightarrow e \tau $ analogous to $H /A \rightarrow \mu \tau$ with the second biggest coefficients $U^{*}_{ki}U_{kj}$ 
can have a non-negligible branching (BMP1).
Moreover, as it is seen from Fig.~\ref{fig:br_AH} for such scenario the $H/A\to bs$ decays have similar branching ratios.  
Some details for such points can be found in Table~\ref{tab:bmps}.

Let us also comment on how the properties of our benchmark points in the $\tB-\cBA$ plane change if one varies the Higgs masses in the range $350-400$ GeV. For BMP1 the splitting between $m_A$ and $m_H$ leads to the problems either with tree-level unitarity or with potential stability. For $m_A=m_H$ in the range $350-400$ GeV the branching ratios of $H$ and $A$ do not change significantly and almost independent of $m_{H^+}$. However, the production cross-sections via gluon fusion drop down by a factor of 2-3. For the case $m_{H^+}<m_{A}$ the mode $H/A \to W^\pm H^\mp$ with off-shell $W^\pm$ starts to contribute. Nevertheless, large mass splitting between neutral and charged higsses, which are required for significant modification of the branching fractions, lead to the problem with potential stability.
For the other two benchmark points the top-pair mode 
dominates the total width of $A$ for $m_A\geq 350$ GeV and of $H$ for $m_H\geq 360$ GeV. 
Due to this, the characteristic branching $H\to hh$ and $H\to \tau \mu$  drop down by a factor of 2-4 when going from $m_H=350$ GeV to $m_H=400$ GeV. 
As for the production of $H/A$ due to the gluon fusion, $gg\to H$ does not change much, while $gg\to A$ is reduced by a factor of 2 for $m_{A/H} = 400$ GeV.

\begin{center}
\begin{table}[h]
\tbl{The benchmark points together with some decay widths and cross sections.}
{\begin{tabular}{|c|c|c|c|c|c|c|c|c|c|c|}
\toprule
$X$  & \rotatebox[origin=b]{90}{$\Br(X \to \tau e) ,~\% $} & \rotatebox[origin=b]{90}{$\Br(X \to \tau \mu ),~\% $} & \rotatebox[origin=b]{90}{$\Br(X \to \tau \tau ),~\% $}& \rotatebox[origin=b]{90}{$\Br(X \to hh ),~\% $} & \rotatebox[origin=b]{90}{$\Br(X \to WW),~\% $} & \rotatebox[origin=b]{90}{$\Br(X \to ZZ),~\% $} & \rotatebox[origin=b]{90}{$\Br(X \to hZ ),~\% $} & \rotatebox[origin=b]{90}{$\Br ( X \to  \gamma \gamma ),~\% $} &  \rotatebox[origin=b]{90}{$\sigma (gg \to X)$,  fb} & \rotatebox[origin=b]{90}{$\Gamma_{tot},~\MeV$}\\
\hline
\multicolumn{11}{|c|}{BMP1: $m_{H}=m_{A}=m_{H^{\pm}}=M=350$ GeV, $(\cBA,\tB) = (0,24)$}\\
\hline
$H$& \textbf{1.39 } & \textbf{27.8} & \textbf{23.9}& 0&  0 & 0 & -- & $5.78 \cdot 10^{-5}$ & 69.8 & 357 \\
\hline
$A$& \textbf{1.34} & \textbf{26.9} & \textbf{23.1}& -- &  -- & -- & 0 & $8.29 \cdot 10^{-5}$ & 182 & 368 \\
\hline
\multicolumn{11}{|c|}{BMP2: $m_{H}=m_{A}=m_{H^{\pm}}=M=350$ GeV, $(\cBA,\tB) = (0.022,6)$}\\
\hline
$H$& 0.649 & \textbf{13.0} & 9.76 & \textbf{30.1} &  9.68 & 4.42 & -- & $2.11 \cdot 10^{-3}$ & 356 & 50.2 \\
\hline
$A$& 0.158 & 3.15 & 2.41 & -- &  -- & -- & \textbf{1.58} & $2.67 \cdot 10^{-3}$ & 1538 & 206 \\
\hline
\multicolumn{11}{|c|}{BMP3: $m_{H}=m_{A}=m_{H^{\pm}}=M=350$ GeV, $(\cBA,\tB) = (0.12,2.51)$ }\\
\hline
$H$& 0.011 & 0.231 & 0.073 & \textbf{58.2}& 23.4 & 10.7 & -- & $1.07 \cdot 10^{-3}$  & 2423 & 619 \\
\hline
$A$& 0.006 & 0.128 & 0.054 & -- & -- & -- & \textbf{8.56} & $\mathbf{2.80 \cdot 10^{-3}}$ & 8450 & 1134 \\
\hline
\end{tabular}
\label{tab:bmps}}
\end{table}
\end{center}

\section{\label{sec:conclusions}Conclusions}

We considered up-type BGL model with tree-level flavor-violating interactions of neutral Higgs bosons.
The BGL symmetry is assumed to be broken softly only by the mass term in the Higgs potential. 
Additional simplification comes from the assumed degeneracy of the neutral CP-odd and charged Higgs bosons.
This model has only four parameters and can be easily confronted with experiment.
Our main goal was to study regions of the model parameter space in which heavy neutral Higgses
can decay into two down-type fermions of different flavour and still escape current experimental bounds 
on direct detection. In our analysis we also took into account recent results on SM Higgs properties together with
theoretical bounds on the Higgs potential parameters. 
We further restrict ourselves by demanding that the masses of heavy Higgses should not deviate much from 
the top-pair threshold $2m_t$ leading to feasible detection in the near future. The allowed regions are studied in $(\tB, \cBA)$ plane and some particular benchmark points with significant $\Br(H/A\to \mu\tau)\lesssim 30\%$ are singled out. 
Our study leads to smaller branchings that stated in Ref.~\refcite{Sher:2016rhh} and demonstrates the crucial role of 
the $t\bar t$ and $c\bar c$ decay modes in the analysis. 
In addition,  we also discuss a few other signatures beside the studied FCNC decay. 
The latter can be used to distinguish different BGL scenarios predicting large $\Br(H/A\to \mu\tau)$ from each other.

In spite of the fact that recent searches \cite{Aad:2016blu,CMS:2016qvi} for the $h\to \mu \tau$ mode do not support  
previous indications \cite{Khachatryan:2015kon,Aad:2015gha} of this decay, we think that the BGL model still deserves attention, since the suppression of FCNC decays of the lightest Higgs 
leads to the enhancement of the corresponding modes for heavy Higgses.

\section*{Acknowledgments}

The authors are grateful to Marc Sher for correspondence regarding Ref.~\refcite{Sher:2016rhh}. 
Financial support from RFBR grant No. 17-02-00872-a is kindly acknowledged.

\end{document}